\newfam\scrfam
\batchmode\font\tenscr=rsfs10 \errorstopmode
\ifx\tenscr\nullfont
	\message{rsfs script font not available. Replacing with calligraphic.}
\else	\font\sevenscr=rsfs7 
	\font\fivescr=rsfs5 
	\skewchar\tenscr='177 \skewchar\sevenscr='177 \skewchar\fivescr='177
	\textfont\scrfam=\tenscr \scriptfont\scrfam=\sevenscr
	\scriptscriptfont\scrfam=\fivescr
	\def\scr{\fam\scrfam}
	\def\cal{\scr}
\fi
\newfam\msbfam
\batchmode\font\twelvemsb=msbm10 scaled\magstep1 \errorstopmode
\ifx\twelvemsb\nullfont\def\Bbb{\bf}
	\message{Blackboard bold not available. Replacing with boldface.}
\else	\catcode`\@=11
	\font\tenmsb=msbm10 \font\sevenmsb=msbm7 \font\fivemsb=msbm5
	\textfont\msbfam=\tenmsb
	\scriptfont\msbfam=\sevenmsb \scriptscriptfont\msbfam=\fivemsb
	\def\Bbb{\relax\expandafter\Bbb@}
	\def\Bbb@#1{{\Bbb@@{#1}}}
	\def\Bbb@@#1{\fam\msbfam\relax#1}
	\catcode`\@=\active
\fi
\font\eightrm=cmr8		\def\xrm{\eightrm}
\font\eightbf=cmbx8		\def\xbf{\eightbf}
\font\eightit=cmti8		\def\xit{\eightit}
\font\eighttt=cmtt8		\def\xtt{\eighttt}
\font\eightcp=cmcsc8
\font\eighti=cmmi8		\def\xold{\eighti}
\font\teni=cmmi10		\def\old{\teni}

\font\tentt=cmtt10
\font\twelverm=cmr12
\font\twelvecp=cmcsc10 scaled\magstep1
\font\fourteencp=cmcsc10 scaled\magstep2
\font\fiverm=cmr5
\font\twelvemath=cmmi12
\font\eightmath=cmmi8

\headline={\ifnum\pageno=1\hfill\else
{\eightcp Cederwall, von Gussich, Nilsson, Westerberg: 
	``The Dirichlet Super-Three-Brane$\ldots$''}
		\dotfill{ }{\old\folio}\fi}
\def\makeheadline{\vbox to 0pt{\vss\noindent\the\headline\break
\hbox to\hsize{\hfill}}
	\vskip2\baselineskip}
\def\makefootline{\ifnum\foottest=1
	\baselineskip=.8cm\line{\the\footline}\global\foottest=0
	\fi
        }
\newcount\foottest
\foottest=0
\def\footnote#1#2{${\,}^#1$\footline={\vtop{\baselineskip=9pt
        \hrule width.5\hsize\hfill\break
        \indent ${}^#1$ \vtop{\hsize=14cm\noindent\xrm #2}}}\foottest=1
        }
\newcount\refcount
\refcount=0
\newwrite\refwrite
\def\ref#1#2{\global\advance\refcount by 1
	\xdef#1{{\old\the\refcount}}
	\ifnum\the\refcount=1
	\immediate\openout\refwrite=\jobname.refs
	\fi
	\immediate\write\refwrite
		{\item{[{\xold\the\refcount}]} #2\hfill\par\vskip-2pt}}
\def\refout{\catcode`\@=11 
	\xrm\immediate\closeout\refwrite
	\vskip2\baselineskip
	{\noindent\twelvecp References}\hfill\vskip\baselineskip
	\parskip=.875\parskip 
	\baselineskip=.8\baselineskip
	\input\jobname.refs 
	\parskip=8\parskip \divide\parskip by 7
	\baselineskip=1.25\baselineskip 
	\catcode`\@=\active\rm}
\newcount\sectioncount
\sectioncount=0
\def\section#1#2{\global\eqcount=0
	\global\advance\sectioncount by 1
	\vskip2\baselineskip\noindent
	\hbox{\twelvecp\the\sectioncount. #2\hfill}\vskip\baselineskip
	\xdef#1{\the\sectioncount}\noindent}
\newcount\appendixcount
\appendixcount=0
\def\appendix#1{\global\aeqcount=0
	\global\advance\appendixcount by 1
	\vskip2\baselineskip\noindent
	\ifnum\the\appendixcount=1
	\hbox{\twelvecp Appendix A: #1\hfill}\vskip\baselineskip\fi
    \ifnum\the\appendixcount=2
	\hbox{\twelvecp Appendix B: #1\hfill}\vskip\baselineskip\fi
    \ifnum\the\appendixcount=3
	\hbox{\twelvecp Appendix B: #1\hfill}\vskip\baselineskip\fi\noindent}
\newcount\eqcount
\eqcount=0
\def\Eqn#1{\global\advance\eqcount by 1
	\xdef#1{{\old\the\sectioncount}.{\old\the\eqcount}}
		\eqno({\oldstyle\the\sectioncount}.
		{\oldstyle\the\eqcount})}
\def\eqn{\global\advance\eqcount by 1
	\eqno({\oldstyle\the\sectioncount}.{\oldstyle\the\eqcount})}
\def\multi{\global\advance\eqcount by 1}
\def\multieq#1#2{\xdef#1{{\old\the\eqcount#2}}
	\eqno{({\oldstyle\the\eqcount#2})}}
\parskip=3.5pt plus .3pt minus .3pt
\baselineskip=14pt plus .1pt minus .05pt
\lineskip=.5pt plus .05pt minus .05pt
\lineskiplimit=.5pt
\abovedisplayskip=18pt plus 2pt minus 2pt
\belowdisplayskip=\abovedisplayskip
\hsize=15cm
\vsize=20.2cm
\hoffset=1cm
\voffset=1.3cm
\def\/{\over}
\def\*{\partial}
\def\a{\alpha}
\def\b{\beta}
\def\c{\gamma}
\def\d{\delta}
\def\e{\varepsilon}

\def\g{\gamma}
\def\k{\kappa}
\def\l{\lambda}

\def\t{\theta}
\def\tb{\bar\theta}
\def\x{\xi}
\def\w{\hskip-.65pt\wedge\hskip-.65pt}
\def\D{\Delta}
\def\F{{\cal F}}
\def\G{\Gamma}
\def\L{{\cal L}}
\def\LL{\Lambda}
\def\P{\Pi}
\def\R{{\Bbb R}}
\def\Ht{\tilde H}
\def\Rt{\tilde R}
\def\V{{\scr V}}
\def\Z{{\Bbb Z}}
\def\punkt{\,\,.}
\def\komma{\,\,,}
\def\.{.\hskip-1pt }
\def\is{\!=\!}
\def\-{\!-\!}
\def\+{\!+\!}
\def\={\!=\!}
\def\>{\!>\!}
\def\kross{\!\times\!}
\def\half{{\lower2.5pt\hbox{\eightrm 1}\/\raise2.5pt\hbox{\eightrm 2}}}
\def\fraction#1{{\lower2.5pt\hbox{\eightrm 1}\/\raise2.5pt\hbox{\eightrm #1}}}
\def\lra{\!\leftrightarrow\!}
\def\tr{\hbox{\rm tr}}
\def\ie{{i.e.,}}
\def\eg{e.g\.}
\def\aka{a.k.a\.}
\def\etal{et al\.}
\def\DBI{\hbox{Dirac--Born--Infeld}}
\def\dbi{{\hbox{\fiverm DBI}}}
\def\wz{{\hbox{\fiverm WZ}}}
\def\det{\hbox{\rm det}}
\def\II{I\hskip-0.6pt I}
%
%
%
%

\null\vskip-1cm
\hbox to\hsize{\hfill G\"oteborg-ITP-96-13}
\hbox to\hsize{\hfill\tt hep-th/9610148}
\hbox to\hsize{\hfill October, 1996}

\vskip4.5cm
\centerline{\fourteencp The Dirichlet Super-Three-Brane} 
\vskip4pt
\centerline{\fourteencp in Ten-Dimensional Type \II B Supergravity}
\vskip\parskip
\centerline{\twelvecp}

\vskip1.2cm
\centerline{\twelverm Martin Cederwall, Alexander von Gussich,} 
\vskip2pt
\centerline{\twelverm Bengt E.W. Nilsson and Anders Westerberg}

\vskip1.2cm
\centerline{\it Institute of Theoretical Physics}
\centerline{\it G\"oteborg University and Chalmers University of Technology }
\centerline{\it S-412 96 G\"oteborg, Sweden}

\vskip.4cm
\catcode`\@=11
\centerline{\tentt tfemc,tfeavg,tfebn,tfeawg@fy.chalmers.se}
\catcode`\@=\active

\vskip2.5cm

\centerline{\bf Abstract}

{\narrower\noindent We give the full supersymmetric 
and $\k$-symmetric action for the Dirichlet three-brane, including its
coupling to background superfields of ten-dimensional 
type \II B supergravity.\smallskip} 

\vfill

\eject

\def\nl{\hfill\break\indent}
\def\nlni{\hfill\break}
\ref\Dai{J.~Dai, R.G.~Leigh and J.~Polchinski, 
	{\xit ``New connections between string theories''}, 
	\nl Mod.~Phys.~Lett. {\xbf A4} ({\xold1989}) {\xold2073}.}   
\ref\Leigh{R.G. Leigh, {\xit ``Dirac--Born--Infeld action from Dirichlet sigma
	model''}, Mod.~Phys.~Lett.~{\xbf A4} ({\xold1989}) {\xold2767}.}
\ref\GreenI{M.B.~Green, {\xit ``Pointlike states for type 2b superstrings},
	Phys.~Lett.~{\xbf B329} ({\xold1994}) {\xold435}
	({\xtt hep-th/9403040}).}  
\ref\Polchinski{J. Polchinski, {\xit ``Dirichlet-branes and Ramond--Ramond 
	charges''},
	\nl Phys.~Rev.~Lett.~{\xbf 75} ({\xold1995}) {\xold4724} 
	({\xtt hep-th/9510017}).}
\ref\Polchinskireview{J.~Polchinski, S.~Chaudhuri and C.V.~Johnson,
	{\xit ``Notes on D-branes''}, {\xtt hep-th/9602052}.}
\ref\PolchinskiWitten{J.~Polchinski and E.~Witten,
	{\xit ``Evidence for heterotic--type I string duality''},
	\nl Nucl.~Phys.~{\xbf B460} ({\xold1996}) {\xold525} 
	({\xtt hep-th/9510069}).}
\ref\WittenIV{E.~Witten, {\xit ``Small instantons in string theory''},
	Nucl.~Phys.~{\xbf B460} ({\xold1996}) {\xold541} 
	({\xtt hep-th/9511030}).}
\ref\Polchinskicoll{J.~Polchinski, {\xit ``String duality --- a colloquium''},
	{\xtt hep-th/9607050}.}
\ref\StromingerVafa{A.~Strominger and C.~Vafa,
	{\xit ``Microscopic origin of the Bekenstein--Hawking entropy''},
	\nl Phys.~Lett.~{\xbf B379} ({\xold1996}) {\xold99}
	({\xtt hep-th/9601029}).}
\ref\Hull{C.M.~Hull and P.K.~Townsend, 
	{\xit ``Unity of superstring dualities''}, 
	\nl Nucl.~Phys.~{\xbf B438} ({\xold1995}) {\xold109} 
	({\xtt hep-th/9410167}).}
\ref\Witten{E.~Witten, {\xit ``String theory dynamics in various dimensions''},
	Nucl.~Phys.~{\xbf B443}
	({\xold1995}) {\xold85} ({\xtt hep-th/9503124}).}
\ref\HullII{C.M.~Hull, {\xit ``String dynamics at strong coupling''},
	Nucl.~Phys.~{\xbf B468} ({\xold1996}) {\xold113} 
	({\xtt hep-th/9512181}).}
\ref\TownsendII{P.K.~Townsend, {\xit ``p-brane democracy''},
	{\xtt hep-th/9507048}.}
\ref\TownsendI{P.K.~Townsend, 
	{\xit ``The eleven-dimensional supermembrane revisited''},
	\nl Phys.~Lett.~{\xbf B350} ({\xold1995}) {\xold184} 
	({\xtt hep-th/9501068}).}
\ref\Schwarz{J.H.~Schwarz, {\xit ``The power of M theory''},
	Phys.~Lett.~{\xbf B367} ({\xold1996}) {\xold97}
	({\xtt hep-th/9510086});\nl
	{\xit ``Lectures on superstring and M theory dualities''},
	{\xtt hep-th/9607201}.}	
\ref\Sen{A.~Sen, {\xit ``Unification of string dualities''},
	{\xtt hep-th/9609176}.}
\ref\Khuri{M.J.~Duff, R.R.~Khuri and J.X.~Lu, {\xit ``String solitons''},
	Phys.~Rep.~{\xbf 259} ({\xold1995}) {\xold213} 
	({\xtt hep-th/9412184}).}
\ref\Duff{M.J.~Duff, {\xit ``Strong/weak coupling duality from the dual
	string''},\nl Nucl.~Phys.~{\xbf B442}
	({\xold1995}) {\xold47} ({\xtt hep-th/9501030}).\vfill\eject}
\ref\Behrndt{K.~Behrndt, E.~Bergshoeff and B.~Jansen,
	{\xit ``Type II duality in six dimensions''},
	\nl Nucl.~Phys.~{\xbf B467} ({\xold1996}) {\xold100} 
	({\xtt hep-th/9512152}).}
\ref\Aharony{O.~Aharony, {\xit ``String theory dualities from M theory''},
	Nucl.~Phys.~{\xbf B476} ({\xold1996}) {\xold470}
	({\xtt hep-th/9604103}).}
\ref\Berkooz{M.~Berkooz, R.G.~Leigh, J.~Polchinski, J.~H. Schwarz, N.~Seiberg
      	and E.~Witten,\nl 
      	{\xit ``Anomalies, dualities, 
		and topology of D=6 N=1 superstring vacua''},\nl
	Nucl.~Phys. {\xbf B475} ({\xold1996}) {\xold115} 
      	({\xtt hep-th/9605184}).}
\ref\BST{E.~Bergshoeff, E.~Sezgin and P.K.~Townsend, 
	\nl {\xit ``Supermembranes and eleven-dimensional supergravity''},
	Phys.~Lett.~{\xbf B189} ({\xold1987}) {\xold75};
	\nl {\xit ``Properties of the eleven-dimensional supermembrane 
	Theory''}, Ann.~Phys. {\xbf 185} ({\xold1988}) {\xold330}.}
\ref\Sezgin{E.~Sezgin, {\xit ``The M algebra''}, {\xtt hep-th/9609086}.}
\ref\Banks{T.~Banks, W.~Fischler, S.H.~Shenker and L.~Susskind,
	\nl{\xit ``M theory as a matrix model: a conjecture''},
	{\xtt hep-th/9610043}.}
\ref\Guven{R.~G\"uven, 
	{\xit ``Black p-brane solutions of D=11 supergravity theory''}, 
	Phys.~Lett.~{\xbf B276} ({\xold1992}) {\xold49}.}
\ref\DuffStelle{M.J.~Duff and K.S.~Stelle,
	{\xit ``Multimembrane solutions of D=11 supergravity''}
	Phys.~Lett.~{\xbf B253} (1991) 113.}
\ref\Blencowe{M.P.~Blencowe and M.J.~Duff, {\xit ``Supermembranes and the 
	signature of space-time''}, \nl Nucl.~Phys.~{\xbf B310}
	({\xold1988}) {\xold387}.}
\ref\Vafa{C. Vafa, {\xit ``Evidence for F-theory''}, 
	Nucl.~Phys.~{\xbf B469} ({\xold1996}) {\xold40}
	({\xtt hep-th/9602022}).}
\ref\Bars{I.~Bars, {\xit ``Algebraic structure of S-theory''},
	({\xtt hep-th/9608061}).}	
\ref\Martinec{D.~Kutasov and E.~Martinec, 
	{\xit New principles for string/membrane unification''},
	\nl Nucl.~Phys.~{\xbf B477} ({\xold1996}) {\xold652} 
	({\xtt hep-th/9602049});
	\nlni D.~Kutasov, E.~Martinec and M.~O'Loughlin, 
	{\xit ``Vacua of M-theory and N=2 strings''},
	\nl Nucl.~Phys.~{\xbf B477} ({\xold1996}) {\xold675} 
	({\xtt hep-th/9603116});
	\nlni E.~Martinec, {\xit ``Geometrical structures of M theory''},
	{\xtt hep-th/9608017}.}
\ref\Ketov{S.V.~Ketov, {\xit ``From N=2 strings to F \& M theory''},
	{\xtt hep-th/9606142}.}
\ref\WittenII{E.~Witten, {\xit ``Twistor-like transform in ten dimensions''},
	Nucl.~Phys.~{\xbf B266} ({\xold1986}) {\xold245}.}
\ref\Grisaru{M.T.~Grisaru, P.~Howe, L.~Mezincescu, B.E.W.~Nilsson and
	P.K.~Townsend, 
	\nl {\xit ``N=2 superstrings in a supergravity background''},
	Phys.~Lett.~{\xbf 162B} ({\xold1985}) {\xold116}.}
\ref\Nilsson{B.E.W.~Nilsson,
	{\xit ``Simple ten-dimensional supergravity in superspace''},
	Nucl.~Phys.~{\xbf B188} ({\xold1981}) {\xold176}.}
\ref\HoweWest{P.S.~Howe and P.C.~West, 
	{\xit ``The complete N=2, d=10 supergravity''},
	Nucl.~Phys.~{\xbf B238} ({\xold1984}) {\xold181}.} 
\ref\BrinkHowe{L.~Brink and P.~Howe, {\xit ``Eleven-dimensional supergravity 
	on the mass-shell in superspace''},
	\nl Phys.~Lett.~{\xbf 91B} ({\xold1980}) {\xold384}.}
\ref\Cremmer{E.~Cremmer and S.~Ferrara, 
	{\xit ``Formulation of eleven-dimensional supergravity 
	in Superspace''},\nl Phys.~Lett.~{\xbf 91B} ({\xold1980}) {\xold61}.}
\ref\BSTII{E.~Bergshoeff, E.~Sezgin and P.K.~Townsend, 
	\nl {\xit ``Super p-branes as gauge theories of volume-preserving
		diffeomorphisms''},
	Ann.~Phys.~{\xbf 199} ({\xold1990}) {\xold340}.}
\ref\DuffLu{M.J.~Duff and J.X.~Lu, 
	{\xit ``Type II p-branes: the brane-scan revisited''}, 
	\nl Nucl.~Phys.~{\xbf B390} ({\xold1993}) {\xold276} 
	({\xtt hep-th/9207060}).}
\ref\HoweSezgin{P.S.~Howe and E.~Sezgin, {\xit ``Superbranes''},
	{\xtt hep-th/9607227}.}
\ref\Strathdee{J.~Strathdee,
	{\xit ``Extended Poincar\'e supersymmetry''},
	Int.~J.~Mod.~Phys.~{\xbf A2} ({\xold1987}) {\xold273}.}  
\ref\Schmidhuber{C. Schmidhuber, {\xit ``D-brane actions''}, 
	Nucl.~Phys.~{\xbf B467} ({\xold1996}) {\xold146}
	({\xtt hep-th/9601003}).}
\ref\Townsend{P.K. Townsend, {\xit ``D-branes from M-branes''}, 
	Phys.~Lett.~{\xbf B373} ({\xold1996}) {\xold68} 
	({\xtt hep-th/9512062}).}
\ref\Tseytlin{A.A.~Tseytlin, {\xit ``Self-duality of Born--Infeld action and 
	Dirichlet 3-brane of Type IIB superstring''},
	\nl Nucl.~Phys.~{\xbf B469} ({\xold1996}) {\xold51}
	({\xtt hep-th/9602064}).}
\ref\BDH{L.~Brink, P.~Di~Vecchia and P.~Howe, 
	{\xit ``A locally supersymmetric and reparametrization invariant 
	action for\nl the spinning string''},
	Phys.~Lett.~\xbf 65B \xrm ({\xold1976}) {\xold471}.}
\ref\Tucker{P.S.~Howe and R.W.~Tucker, 
	{\xit ``A locally supersymmetric and reparametrization invariant
	action for a \nl spinning membrane''}, 
	J.~Phys.~{\xbf A10} ({\xold1977}) L{\xold155}.\vfill\eject}
\ref\Callanetal{C.G. Callan, C. Lovelace, C.R. Nappi and S.A. Yost,
	{\xit ``String loop corrections to beta functions''},
	\nl Nucl.~Phys. {\xbf B288} ({\xold1987}) {\xold525}.}
\ref\CGMNW{M.~Cederwall, A.~von~Gussich, A.~Mikovi\'c, B.E.W.~Nilsson
	and A.~Westerberg, \nl{\xit ``On the \DBI\ action for D-branes''},
	Phys.~Lett.~B (to appear), {\xtt hep-th/9606173}, .}
\ref\BdRO{E.~Bergshoeff, M.~de Roo and T.~Ort\'{\i}n,
	{\xit ``The eleven-dimensional five-brane''},
	{\xtt hep-th/9606118}.}
\ref\GHT{M.B.~Green, C.M.~Hull and P.K.~Townsend, \nl {\xit ``D-brane 
	Wess--Zumino actions, T-duality and the cosmological constant''},
	{\xtt hep-th/9604119}.}
\ref\Achucarro{A.~Ach\'ucarro, J.M.~Evans, P.K.~Townsend and D.L.~Wiltshire,
	{\xit ``Super p-branes''},
	\nl Phys.~Lett.~{\xbf B198} ({\xold1987}) {\xold441}.} 
\ref\WessBagger{J.~Wess and J.~Bagger, 
	{\xit ``Supersymmetry and supergravity''},
	Princeton University Press (Princeton NJ, 1992).}
\ref\Douglas{M.~Douglas, {\xit ``Branes within branes''},
	{\xtt hep-th/9512077}.}
\ref\CFNW{M.~Cederwall, G.~Ferretti, B.E.W.~Nilsson and A.~Westerberg, 
	{\xit ``Higher-dimensional loop algebras, \nl non-abelian extensions 
	and p-branes''},
	Nucl.~Phys.~{\xbf B424} ({\xold1994}) {\xold97} 
	({\xtt hep-th/9401027}).}
\ref\Bergshoeff{E.~Bergshoeff, C.M.~Hull and T.~Ort\'{\i}n,
	{\xit ``Duality in the type-II superstring effective action''},
	\nl Nucl.~Phys.~{\xbf B452} ({\xold1995}) {\xold547}
	({\xtt hep-th/9504081}).}
\ref\Gutperle{M.B.~Green and M.~Gutperle, {\xit ``Comments on 3-branes''},
	Phys.~Lett.~\xbf B377 \xrm ({\xold1996}) {\xold28}
	({\xtt hep-th/9602077}).}
\ref\Karlhede{A.~Karlhede, U.~Lindstr\"om, M.~Ro\v cek and G.~Theodoridis,
	{\xit ``Supersymmetric nonlinear Maxwell theories and 
		\nl the string effective action''},
	Nucl.~Phys.~{\xbf B294} ({\xold1987}) {\xold498}.}
\ref\WittenIII{E.~Witten, {\xit ``Bound states of strings and p-branes''},
	Nucl.~Phys.~{\xbf B460} ({\xold1996}) 335 ({\xtt hep-th/9510135}).}

\section\intro{Introduction}The recent incorporation of 
Dirichlet $p$-branes in string theory has led
to major advances in our understanding of the non-perturbative aspects of
the theory [\Dai,\Leigh,\GreenI,\Polchinski,\Polchinskireview].
This includes issues like the r\^oles played by solitonic states in
non-perturbative string theories and results concerning string
duality [\PolchinskiWitten,\WittenIV,\Polchinskicoll] 
as well as fundamental explanations of results known from semi-classical
gravity [\StromingerVafa].
An important aspect is the fascinating r\^ole played by standard
supergravity theories in uncovering deep connections between different
perturbative sectors of the full non-perturbative theory
[\Hull,\Witten,\HullII]. It is now clear that
eleven-dimensional supergravity, which previously seemed completely 
disconnected from string theory, plays a rather central, unifying r\^ole 
in the full theory [\Witten], \eg\ when discussing a higher-dimensional origin
of $p$-brane solutions [\TownsendII]  and  various kinds of duality symmetries 
appearing in lower-dimensional ($D\!\!\leq\!\!10$) theories 
[\Witten,\TownsendI-
\Berkooz\ and references therein]
(we do not aim at making the reference list complete).

In the same spirit as one obtains supergravity theories from strings
in ten dimensions and below, it was argued some time ago that one 
should seek to explain the existence of eleven-dimensional supergravity
in terms of the enigmatic quantized supermembrane [\BST], or,
as suggested more recently, in terms of the even less understood M-theory
[\Witten,\Schwarz,\Sen,\Sezgin,\Banks], of which the membrane might be 
a low-energy manifestation. That such a connection might exist is indicated
by the fact that $p$=2 (and $p$=5) branes [\Guven,\DuffStelle] occur as
solutions to the $D\=11$ supergravity field equations and by the relation
between $\k$-symmetry and the constraints on the $D\=11$ supergeometry
that implies the field equations.

However, some interesting chiral string theories and their $p$-brane solutions 
fall outside this scheme, as emphasized in [\HullII], 
and instead hint at the existence of new supergravity-like 
theories in yet higher dimensions [\Blencowe], F-theory [\Vafa] 
(or Y-theory [\HullII]) in twelve
dimensions or fundamentally new theories in thirteen [\Bars,\Martinec,\Ketov] 
dimensions or above,
often with non-standard signatures [\Vafa,\Martinec,\Ketov]. 
The search for these exotic theories would probably benefit from a more 
detailed understanding of the less exotic objects known to exist in 
eleven dimensions and below. 
One particularly interesting piece of new information
would be the construction of 
$\k$-symmetric covariant actions which involve vector or antisymmetric
tensor supermultiplets in a fundamental way.

Higher-dimensional extended objects of various kinds have turned out to
be extremely useful in many of the recent discussions of these matters
(see \eg\ refs\.\ [\Polchinskireview,\Hull]), 
in spite of the fact that they are in many cases rather poorly understood. 
For example, in ten and eleven dimensions full knowledge of the covariant
super- and $\k$-symmetric action including background couplings has been
obtained only for the string ($p$=1) [\WittenII,\Grisaru] in $D\=10$, 
$N\=1$ supergravity 
[\Nilsson] and type IIB supergravity [\HoweWest], 
for the membrane ($p$=2) [\BST] in $D\=11$ [\BrinkHowe,\Cremmer]
and for the 5-brane in $D\=10$ [\BSTII]. 
Although a more complete picture 
[\DuffLu,\HoweSezgin] of possible world-volume 
supermultiplets [\Strathdee] is now emerging,
none of the actions based on vector or antisymmetric tensor supermultiplets 
have yet been constructed. A case that falls in between is the membrane in
type IIA supergravity which has Ramond--Ramond (RR) charges and hence is
a D-brane with $p$=2. We are here referring to the fact that the action for 
this type IIA membrane can be obtained by dualizing [\DuffLu,
\Schmidhuber,\Townsend,\Tseytlin] the eleventh coordinate for
the membrane in $D$=11 to a vector, thereby generating an action
containing an abelian world-volume vector field. 
Using the formulation of Bergshoeff \etal\ [\BST], 
the action derived in this way by Townsend 
[\Townsend] is of Brink--DiVecchia--Howe--Tucker (BDHT, \aka\ Polyakov) 
type with a non-propagating world-volume metric [\BDH,\Tucker]. 
Elimination of this auxiliary metric by means of 
its field equation leads to a \DBI\ (DBI) action. 
That such an action must appear is known from open string calculations in the 
case of constant field strength for the Born--Infeld vector field, at least 
for the bosonic part of the action [\Leigh,\Callanetal]. 
In fact, a fully covariant target space supersymmetric and 
world-volume $\k$-symmetric DBI action has not yet been presented in the 
literature. 

It is the purpose of this paper to derive such an action for
the case of the Dirichlet 3-brane in type IIB string theory. In this case
the answer can not be deduced by dualizing any previously known
action. Instead, our derivation
starts from the bosonic DBI action and knowledge about the structure of  
the necessary Wess--Zumino terms obtained from the superspace formulation
of type IIB supergravity. 
As is the case for the membrane propagating in the background
of eleven-dimensional supergravity, the action for the D3-brane in type IIB
superspace is only $\k$-symmetric provided the background fields 
satisfy the proper on-shell constraints (see ref\.\ [\BST]). 
Similar results in the case of type IIB were obtained 
some years ago by Grisaru \etal\ [\Grisaru], 
who coupled the type IIB supergravity to
the string in the NS-NS sector. Note however that in this case one does not
probe the constraints of the field strengths coming from the RR sector.
In the case of the D3-brane considered here on the other hand, all
constraints necessary to derive the field equations of type IIB supergravity
[\HoweWest] are involved in establishing the $\k$-invariance of the action.

The plan of the paper is as follows.
In section {\old2} we explain our notation and conventions, and present the 
generalization of the 
bosonic action to a $\k$-symmetric action in flat type IIB superspace. 
Section {\old3} is devoted to the construction of the couplings to 
the on-shell type IIB background superfields,
which as mentioned above involves all the field strengths of the theory,
\ie\ both the ones that originate in the NS-NS sector and the ones from 
the RR sector. 
For the bosonic part of the action the exact relation between
the BDHT version and the DBI version is known from the work of 
ref\.\ [\CGMNW]. A summary of our results including a discussion of the
$\k$-symmetric BDHT-type extension of ref\.\ [\CGMNW] 
and some final comments are collected in section {\old4}.

\section\Conventions{The {\twelvemath\char'24}-symmetric {\twelvemath p}=3 
action in flat IIB superspace}We
will start the construction by considering the  bosonic DBI action
$$
\L_{\dbi}=-\sqrt{-\det(g_{ij}+F_{ij})}\komma\eqn
$$
where $g_{ij}$ is the pullback to the four-dimensional world-volume
of the flat ten-dimensional target space metric $\eta_{ab}$
and $F_{ij}$ is the field strength of the world-volume vector field appearing
in all D-brane actions generated from open strings. (The standard 
factor of ${\a'\/2\pi}$ in front of $F_{ij}$ 
is set equal to one in this paper.) The pullback is 
written in terms of $\partial_iX^a$ where $X^a$ are coordinates 
in the flat ten-dimensional bosonic target space. As a  first step we now 
generalize the corresponding world-volume form $dX^a$ to a globally 
supersymmetric form by introducing the
type IIB fermionic coordinates $\t^{\a}$. Since we will
throughout this paper use real sixteen-component Majorana--Weyl 
spinors and the
corresponding $\g$-matrices we must remember that type IIB spinors have an 
extra two-dimensional SO$(2)$ index and that indices $\a,\b,\ldots$ are thus 
to be viewed
as composite indices representing the tensor product of a Majorana--Weyl 
index and 
an SO$(2)$ index. The $\g$-matrices are extended in a similar fashion and
all these quantities will as a consequence remain real. The $\g$-matrices
anticommute to $2\eta_{ab}$ (with a suppressed ${\d_{\a}}^{\b}$) 
which we choose to be 
a mostly positive flat metric.

The supersymmetric world-volume variables (forms) for the case where there 
are no background fields are
$$
\eqalign{&d\t^{\a}\komma\cr
        &\P^a=dX^a+id\tb\g^a \t\komma\cr
	&\F=dA-idX_a\w(d\tb\g^aK\t)+\half(d\tb\g_a\t)\w(d\tb\g^aK\t)\punkt\cr}
\eqn
$$
The expression for $\F$ contains the matrix $K$, which is a real 
2$\times$2-matrix with $K^2\=1$. It acts on the SO$(2)$ indices
mentioned above and together with $I$ and $J$ 
($I^2\=-1$, $J^2\=K^2\=1$, $IJ\=K$ and $I,J,K$ anticommuting) they  behave
as gamma-matrices (generators) for SL$(2;\R)\cong\,$Spin$(1,2)$, 
or, equivalently, as the imaginary split quaternionic units.
A convenient basis (although unnecessary for any calculations) is
$$
I=\left[\matrix{0&1\cr-1&0\cr}\right]\komma\qquad
J=\left[\matrix{0&1\cr1&0\cr}\right]\komma\qquad
K=\left[\matrix{1&0\cr0&-1\cr}\right]\komma\eqn
$$
{}from which one possible correspondence with a complex spinor formalism
may be read off: $I\psi\lra-i\psi$, $J\psi\lra i\psi^*$, $K\psi\lra\psi^*$.
Concerning the conventions for ordinary forms, we have here adopted the 
ones that are standard in superspace 
differential geometry, namely $\F\={1\/2}d\x^j\w d\x^i\F_{ij}$ 
(note the order of the indices). As will be 
discussed in the next section, these globally supersymmetric extensions 
arise naturally when taking the flat superspace limit of the
D3-brane in a general background. 
The reason we present the flat case first is that it clarifies how the 
$\k$-symmetry works without any need for knowledge about the rather 
complicated superspace formulation of type IIB supergravity. 

The rigid supersymmetry transformations under which 
the above forms are invariant are
$$
\eqalign{&\d_\e\t^{\a}=\e^{\a}\komma\phantom{\half}\cr
	&\d_\e X^a=-i(\bar\t\g^a\e)\komma\phantom{\half}\cr
	&\d_\e A=idX_a(\bar\t\g^aK\e)-\fraction{6}(d\tb\g_a\t)(\bar\t\g^aK\e) 
		-\fraction{6}(d\tb\g_aK\t)(\bar\t\g^a\e)\komma\cr}
\Eqn\epsilonvariations
$$ 
where $\e$ is a constant target space IIB spinor and a world-volume scalar.
It is a priori clear that $A$ must transform under supersymmetry, since
it contains two of the eight physical bosonic degrees of freedom matching 
the fermionic ones. We will comment further on this issue in section {\old3}.
To prove the invariance of $\F$ one must make use of the ``cyclic
Fierz identity''
$$
(\bar A\g_aB)(\bar C\g^aD)=-\half(\bar A\g_ae_ID)(\bar B\g^ae_IC)
		-\half(\bar A\g_ae_IC)(\bar B\g^ae_ID)\komma\Eqn\fierzid
$$
where $e_I\=1,I,J,K$ and summation over the index $I$ is understood.
 Note that $d\t$
anticommutes with $\t$ but commutes with another $d\t$. By writing
$A=\a\psi$, with $\a$ an odd parameter, the above identity, which is 
valid for bosonic spinors $A,B,C,D$, can be used to
derive Fierz identities for any kind of spinors. Equation (\fierzid) is
the only Fierz identity needed throughout this paper. By using it twice,
one finds, \eg,
$$
(\g_a)_{(\a\b}(\g^a)_{\g)\d}=-(\g_aJ)_{(\a\b}(\g^aJ)_{\g)\d}
      =-(\g_aK)_{(\a\b}(\g^aK)_{\g)\d}\komma\eqn
$$
identities which are used repeatedly below in calculations concerning 
$\k$-symmetry and Bianchi identities.

Turning to the $\k$-symmetry, we know from previous experience that the 
$\k$-transformations of $\t^\a$ and $X^a$ 
differ from the $\e$-transformations above 
only by a sign in
the transformation for $X^a$ (this fact will find its natural explanation in
section {\old3} --- a $\k$-transformation is rather 
half a covariant derivative,
and commutes with supersymmetry).
The $\k$-transformation of $A$ is completely fixed once one demands that
$\k$-transformations are supersymmetric, \ie\ that $\d_\e$ and $\d_\k$ commute.
To determine the overall constant, one makes the inspired guess that 
$\d_{\k} g_{ij}$ and $\d_{\k} \F_{ij}$ should end up having a 
similar structure.
Using the transformations
$$
\eqalign{&\d_\k\t^{\a}=\k^{\a}\komma\phantom{\half}\cr
	&\d_\k X^a=i(\bar\t\g^a\k)\komma\phantom{\half}\cr
	&\d_\k A=-idX_a(\bar\t\g^aK\k)+\half(d\tb\g_a\t)(\tb\g^aK\k) 
		+\half(d\tb\g_aK\t)(\tb\g^a\k)\komma\cr}\eqn
$$ 
we find
$$
\eqalign{&\d_\k\P^a=2i(d\bar\t\g^a\k)\komma\cr
	&\d_\k\F=-2i\P_a\w(d\bar\t\g^aK\k)\punkt\cr}\eqn
$$

In order for the bosonic and fermionic fields in the 
world-volume action to have
equal numbers of degrees of freedom we must as usual project out half of the
components of the local space-time spinor parameter $\k$. 
This is accomplished by imposing $\k\={1\/2}(1\+\G)\k$, where
$$
\G={\e^{ijkl}\/\sqrt{-\det(g+\F)}}\left(\fraction{24}\g_{ijkl}I
	-\fraction{4}\F_{ij}\g_{kl}J+\fraction{8}\F_{ij}\F_{kl}I\right)\eqn
$$
satisfies $\G^2\=1$. Here, $\g_{i_1\ldots i_n}$ is the pullback to the 
world-volume of the antisymmetrized (weight one) product of $n$ $D\=10$
$\g$-matrices, and thus $\g_{(i}\g_{j)}=g_{ij}$.
In proving that the matrix $\G$ squares to one,
it is convenient to rewrite the DBI lagrangian using
$$
\det(g_{ij}+\F_{ij})=\det(g_{ij})
\bigl(1 -\half \tr(g^{-1}\F)^2\bigr)+\det(\F_{ij})
\komma\eqn
$$
and make use of identities like
$$
\eqalign{
&\det\F=(\fraction{8}\e^{ijkl}\F_{ij}\F_{kl})^2\komma\cr
&(g^{-1}\F)^4=\half(g^{-1}\F)^2\tr(g^{-1}\F)^2-\det(g^{-1}\F)\komma\cr}
\eqn
$$
valid for antisymmetric matrices $\F_{ij}$. Given an antisymmetric tensor,
the form of $\G$ is unique up to the choice for the matrix $J$, which in
turn depends on the choice of the matrix $K$ in the transformations.
A perhaps trivial remark is that just by trying to construct a projection
matrix for $p\=3$, one is forced to introduce a pair of Majorana--Weyl spinors
of equal chirality. The question of chirality is related
to the sign of the square of $\e^{i_1\ldots i_{p+1}}\g_{i_1\ldots i_{p+1}}$,
which is $(-1)^{{1\/2}p(p+1)+1}$. Already this observation tells much about
the structure of supersymmetric branes of diverse dimensionalities. The
construction of $\G$ is of course a key ingredient in establishing 
$\k$-symmetry, and therefore in the entire construction of the supersymmetric
brane action.

After a somewhat tedious calculation one arrives at the following result
for the $\k$-variation of the DBI lagrangian:
$$
\d_\k\left(-\sqrt{-\det(g+\F)}\right)
	=\e^{ijkl}\left(
	-{\lower2.5pt\hbox{\eightmath i}\/\raise2.5pt\hbox{\eightrm 3}}
	(\*_i\bar\t\g_{jkl}I\k)+
	i\F_{ij}(\*_k\bar\t\g_lJ\k)\right)\punkt
	\Eqn\lzerovariation
$$
The computations are simplified somewhat if one instead of
inserting the projected parameter
${1\/2}(1\+\G)\k$ just uses $\G\k$. Note that the integral of the right hand 
side can be nicely
written as the integral of a world-volume four-form,
$$
\d_\k I_\dbi=\int\Bigl(\,2i(d\tb\g_{(3)}I\k)-2i\F(d\tb\g_{(1)}J\k)\,\Bigr)
\komma\Eqn\dbivariation
$$
where 
$\g_{(n)}\={1\/n!}d\x^{i_n}\w\,\ldots\,\w d\x^{i_1}\g_{i_1}\ldots\g_{i_n}$. 
The result that the $\k$-variation of the DBI action gives the integral 
of a differential form is of course a necessary condition for the 
existence of an invariant action built out of just a DBI term and
a Wess--Zumino (WZ) term.

In order to find the WZ part of the action, it is in principle
possible to make a general ansatz and determine the exact expression by
demanding cancellation with (\dbivariation). 
However, a better alternative would be to
first consider its general superspace
structure discussed in refs\.\ [\BdRO,\GHT], and
make use of the already solved type IIB superspace Bianchi 
identities [\HoweWest] to obtain the exact expression. The flat limit
is then expected to produce the same answer as the more direct approach
alluded to above. Since  
this will be carried out in detail in section {\old3}, where
we will discuss the coupling of the D3-brane
to a general type IIB supergravity background, we end this section by 
simply giving the expression for the WZ action whose variation will cancel
the one from the DBI term:
$$
I_{\wz}=\int e^{\F}C=\int (C_{(4)}+\F\wedge C_{(2)})\komma\Eqn\flataction
$$
where
$$
C_{ab}=0\komma\qquad C_{a\b}=i(\g_a J\t)_\b\komma\qquad 
C_{\a\b}=(\g^a\t)_\a(\g_aJ\t)_\b
\komma\eqn
$$
and
$$
\eqalign{&C_{abcd}=0\komma\cr
	&C_{abc\d}=-i(\g_{abc}I\t)_\d\komma\cr
	&C_{ab\c\d}=-(\g^c\t)_\c(\g_{abc}I\t)_\d
		-2(\g_aJ\t)_\c(\g_bK\t)_\d\komma\cr
	&C_{a\b\c\d}=i(\g^b\t)_\b(\g^c\t)_\c(\g_{abc}I\t)_\d
	-2i(\g^b\t)_\b(\g_bJ\t)_\c(\g_aK\t)_\d
	-i(\g^b\t)_\b(\g_bK\t)_\c(\g_aJ\t)_\d\komma\cr
	&C_{\a\b\c\d}=(\g^a\t)_\a(\g^b\t)_\b(\g^c\t)_\c(\g_{abc}I\t)_\d
	-3(\g^a\t)_\a(\g_aJ\t)_\b(\g^b\t)_\c(\g_bK\t)_\d\komma\cr}\eqn
$$
where (anti-)symmetrization is understood.
In the following section, the various $\t$-extended
quantities introduced here will receive a proper 
explanation in terms of the flat limit
of the superspace formulation of the target space type IIB supergravity theory.

When one considers an ordinary $p$-brane, only the first term in equation
(\flataction) is present, $I^{(p)}_\wz=\int C_{(p+1)}$, and the crucial 
$\g$-matrix identity, yielding the ``brane scan''[\Achucarro], is
$$
(\g^{a_1\ldots a_{p-1}a})_{(\a\b}(\g_a)_{\g\d)}=0\punkt\eqn
$$
The corresponding identity here, which can be seen as responsible for the
existence of the D3-brane, and which is also crucial for the Bianchi 
identities of the following section, is
$$
(\g^{abc}I)_{(\a\b}(\g_c)_{\g\d)}-2(\g^{[a}J)_{(\a\b}(\g^{b]}K)_{\g\d)}=0
\punkt\eqn
$$
A proper classification of similar identities in different dimensionalities
should provide a ``D-brane scan'' [\DuffLu].

\section\general{Coupling to a general IIB background}As is well known 
(see \eg\ ref\.\ [\HoweWest]), the field theory of ten-dimensional 
type IIB supergravity contains in the bosonic sector 
the metric $g_{mn}$, a complex scalar $\l\=C_{(0)}\+ie^{-\phi}$, a complex
two-form $a_{(2)}\=C_{(2)}\+iB_{(2)}$ and a self-dual real four-form $C_{(4)}$,
and in the fermionic sector the complex spinor $\LL_{\a}$ 
and the Rarita--Schwinger
field ${\psi_m}^{\a}$. Bosonic fields originating in the RR sector of the 
string are here denoted as $C_{(n)}$. 
The indices $m,n,\ldots$ are ten-dimensional world indices
which will be combined with $\mu,\nu,\ldots$ to ten-dimensional super-world 
indices $M,N,\ldots$. The superspace vielbein is then defined by
$$
E^A=dZ^M {E_M}^{A}\komma\eqn
$$
where $dZ^M$ are the superspace differentials $(dX^m,d\t^{\mu})$. 
Our superspace differential geometry conventions are those of Wess and Bagger
[\WessBagger].

Since the action for a D$p$-brane in general involves a DBI term with
couplings only to NS-NS fields  while the RR couplings appear exclusively
in the WZ term, it is not
possible to work with the above complex combinations of NS-NS and RR fields.
It turns out to be necessary to keep also the spinors real, as we did in the
previous section. It will even turn out that the natural choice of
field strength does not correspond to the real and imaginary part of
a complex field.
As a consequence, some formul\ae\ below may look somewhat unfamiliar to a 
reader used to the previous superspace treatments of type IIB 
supergravity [\HoweWest].
Real Majorana--Weyl spinors are here written in terms of composite 
indices $\a,\b,\ldots$ in tangent space or $\mu,\nu,\ldots$ 
in curved superspace.

In superspace all ordinary component fields are introduced as first components
of their corresponding superfields. From the gauge fields one then constructs
the super-field strengths and derives the super-Bianchi identities. 
In order to reduce
the enormous field content of these superfields down to the on-shell
content one introduces constraints on some of the components of the super-field
strengths. When these constraints are inserted in the Bianchi 
identities, the latter cease to be 
identities, and if the constraints are properly chosen the equations so
obtained are just the supergravity equations of motion. For the membrane
in eleven-dimensional supergravity it turns out that it is exactly these
on-shell 
constraints (modulo different choices of the conventional ones) that must be
imposed in order for the membrane action to be $\k$-symmetric [\BST].
We will now show that the same phenomenon is at work here.

The action is $I=I_{\dbi}+I_{\wz}$, where 
$$
I_{\dbi}=-\int d^4\xi\sqrt{-\det(\hat g+e^{-{\phi\/2}}\hat\F)}
\Eqn\superDBI
$$
and 
$$
I_{\wz}=\int e^{\hat\F}\hat C \punkt\Eqn\superWZ
$$
This expression formally agrees with refs\.\ [\Douglas,\GHT,\Tseytlin,\BdRO],
with all bosonic fields replaced by the corresponding superfields.
Here $\hat g$ refers to the pullback of the ten-dimensional flat metric
$\eta_{ab}$, \ie\ 
$$
\hat g_{ij}={E_i}^a {E_j}^b \eta_{ab}\komma\eqn
$$
which in an arbitrary background involves the vielbein through
${E_i}^A\=(\*_i Z^M){E_M}^A$. 
The use of hats indicates coupling to a general type IIB superspace 
background. For instance, in the NS-NS sector the
superfield $B_{M\!N}$ enters through 
$\hat\F\={1\/2}d\xi^j\!\w d\xi^i(F_{ij}\-\hat B_{ij})$
where $\hat B_{ij}\= {E_j}^B {E_i}^A B_{AB}$. 
In the WZ term the pullback of the
RR potential $\hat C$ is defined in a similar fashion. Note that we have 
adopted the definition of $\hat C$ introduced in ref\.\ [\GHT]
where this RR potential is considered as a sum of forms
of different degrees. In the case of type IIB they are all of even degree.
In fact, such a sum was already discussed by
Cederwall et al\.\ [\CFNW] in the investigation of $p$-brane algebras and 
their general extensions.

The advantage of equation (\superDBI) over the form 
$\L_{\dbi}\=-e^{-\phi}(-\det(\hat g_s\+\hat\F))^{1\/2}$, 
$\,\hat g\=e^{-{\phi\/2}}\hat g_s$, is that 
the metric $\hat g$ is the SL$(2;\R)$-invariant Einstein metric, whereas
$\hat g_s$ is the string metric and transforms under SL$(2;\R)$.
For $\hat g$ (and all associated superfields) we can use the conventional
supergravity constraints [\HoweWest], while the torsion corresponding to
the vielbein of $\hat g_s$ necessarily contains explicit dilaton factors.
It is a property peculiar to $p\is3$, and essential for its duality
properties, that the naked Einstein metric occurs in the action 
[\Bergshoeff,\Tseytlin,\Gutperle].

When written in this way, the D3-brane action is manifestly invariant
under type IIB superspace reparametrizations. 
One might wonder how this is reconciled with the coupling to open strings
where the boundary conditions break $N\=2$ to $N\=1$. The answer is identical
to the situation for the gauge invariance of an antisymmetric tensor field
coupling to the bulk of an open string. Na\"\i vely, the gauge invariance is
destroyed by the presence of a boundary, but when the boundary is a
D-brane it is restored by the coupling of the endpoints to the vector 
potential $A$ on the world-volume. The gauge transformation $\d B\=d\lambda$
is accompanied by $\d A\=\lambda$, so that the combination 
$\hat\F\=F\-\hat B_{(2)}$ is invariant. For $N\=2$ supersymmetry, an
exactly parallel mechanism is at work, and this is the concrete interpretation
of the third equation in (\epsilonvariations), which is exactly what
is necessary to keep $\F$ invariant in the flat case [\Witten,\Townsend].
In fact, once the extra terms in the modified supersymmetric field strength
are understood as the pullback of the superfield $B$, the just mentioned
gauge transformation can be used to move the transformations
freely between $A$ and $B$, and in particular to keep $A$ inert under
supersymmetry and $\k$-symmetry. 

We now proceed to the issue of $\k$-symmetry.
Contrary to the situation in section {\old2}, we now have full knowledge of the
WZ term from the outset. This gives us the possibility to perform the 
calculations in a slightly different manner. By first varying
the WZ term with parameter $\G\k$ one finds that the calculations simplify
quite a bit. The reason for this is that one directly obtains
the factor $\L_{\dbi}^{-1}$, which also follows from the variation of the
DBI term if written in the form
$$\eqalign{
\d_{\k}\L_{\dbi}&=-\fraction{12}{\e^{ijkl}\e^{i'j'k'l'}\/\L_{\dbi}}
(\hat g_{ii'}+e^{-{\phi\/2}}\hat\F_{ii'})
(\hat g_{jj'}+e^{-{\phi\/2}}\hat\F_{jj'})\cr
&\qquad\times(\hat g_{kk'}+e^{-{\phi\/2}}\hat\F_{kk'})
	(\d_{\k}\hat g_{ll'}+e^{-{\phi\/2}}\d_{\k}\hat\F_{ll'}-
\half\d_{\k}\phi e^{-{\phi\/2}}\hat \F_{ll'})\punkt\cr}\Eqn\termbyterm
$$
Consequently, there will be no need for any of the matrix identities 
essential for proving $\G^2\=1$, and we find that the comparison of the 
terms coming from the variation of the DBI and WZ parts, respectively, 
is rather straightforward.

In order to derive the variations to be inserted in equation (\termbyterm) we 
need 
the variations of the superspace coordinates. These can be written in 
the following form [\BST]:
$$
(\d_{\k}Z^M){E_M}^a=0\komma\qquad(\d_{\k}Z^M){E_M}^{\a}=\k^{\a} 
\punkt\Eqn\zkappas
$$
We then find that
$$\eqalign{&\d_{\k}\hat g_{ij}=2{E_{(i}}^a{E_{j)}}^B \k^{\a}{T_{\a Ba}}
		\komma\cr
	&\d_{\k}\hat\F_{ij}=-{E_j}^B{E_i}^A\k^{\a}H_{\a AB}\komma\cr
	&\d_{\k}\phi=\k^{\a}\*_{\a}\phi\punkt\cr}\eqn
$$
To be explicit, the transformation of $\hat B$ induced by (\zkappas) is
$\d_\k\hat B=\L_\k\hat B\equiv(d\,i_\k+i_\k d)\hat B=d\,i_\k\hat B+
i_\k\hat H$, where the first term is cancelled in $\d_\k\hat\F$ by
the gauge transformation $\d A\=i_\k\hat B$ (see the discussion above), 
leaving $\d_\k\hat\F=-i_\k\hat H$.

At this point one needs to
solve the superspace Bianchi identities [\HoweWest] to verify that
the for us relevant on-shell constraints (those with dimension 0 and ${1\/2}$)
are 
$$\eqalign{
&{T_{\a\b}}^c=2i(\g^c)_{\a\b}\komma
	\phantom{H_{\a\b c}=2ie^{{\phi\/2}}(\g_c K)_{\a\b}}\hskip-2cm
	{T_{\a b}}^c=0\komma\cr
	&{T_{\a\b}}^\c=-{(J)_{(\a}}^\c(J\LL)_{\b)}
		+{(K)_{(\a}}^\c(K\LL)_{\b)}\cr
		&\phantom{{T_{\a\b}}^\c=}+\half(\g_aJ)_{\a\b}(\g^aJ\LL)^\c
		-\half(\g_aK)_{\a\b}(\g^aK\LL)^\c\komma\cr
	&H_{\a\b c}=2ie^{{\phi\/2}}(\g_c K)_{\a\b}\komma
	\phantom{{T_{\a\b}}^c=2i(\g^c)_{\a\b}}\hskip-2cm	
		H_{\a bc}=-e^{{\phi\/2}}(\g_{bc}K \LL)_{\a}\komma\cr
	&\*_{\a}\phi =2\LL_{\a}\punkt\cr}
	\Eqn\nsnsconfig
$$
(We will demonstrate in a little while that this configuration 
is the correct one.)
Thus, we can write the variation in the following form
$$\eqalign{&\d_{\k}\hat g_{ij}=4i(\bar E_{(i}\g_{j)}\k)\komma\cr
	&\d_{\k}\hat \F_{ij}=-4ie^{\phi\/2}(\bar E_{[i}\g_{j]}K\k)
		+e^{\phi\/2}(\bar\LL\g_{ij}K\k)\komma\cr
	&\d_{\k}\phi=2\bar\k \LL\punkt\cr}\eqn
$$

Modulo boundary terms, the variation of the WZ term gives
$$
\d_{\k}I_{\wz}=\int\left((i_{\k}\hat R)_{(4)}+\hat\F\wedge(i_{\k}\hat R)_{(2)} 
+\half \hat\F\wedge\hat\F \wedge(i_{\k}\hat R)_{(0)}\right)\komma\eqn
$$
where the hatted field strengths $\hat R$ are pullbacks of 
$(i_{\k} R)_{(n)}={1\/n!}E^{A_n}\w\,\ldots\,\w E^{A_1}\k^\a 
R_{\a A_1\ldots A_n}$. Thus we find that only
super-field strength components with at least one spinor index are relevant.

The formal sum of the curvature forms can be written 
$R\=e^Bd(e^{-B}C)\=dC\-H\w C$, \ie
$$
\eqalign{&R_{(5)}=dC_{(4)}-H\w C_{(2)}\komma\cr
	&R_{(3)}=dC_{(2)}-H\w C_{(0)}\komma\cr
	&R_{(1)}=dC_{(0)}\komma\cr}\eqn
$$
so the Bianchi identities read $d(e^{-B}R)\=0$, \ie
$$
\eqalign{&dR_{(5)}+H\w R_{(3)}=0\komma\cr
	&dR_{(3)}+H\w R_{(1)}=0\komma\cr
	&dR_{(1)}=0\punkt\cr}\Eqn\bianchiidentities
$$
In the real basis employed here, the superspace constraints
that enter into the variation of the WZ term read (the exact relation to
the constraints given by Howe and West in ref.~[\HoweWest] will soon be
established)
$$
\eqalign{&R_{abc\a\b}=2i(\g_{abc}I)_{\a\b}\komma
	\phantom{R_{a\b\c}=-2ie^{-{\phi\/2}}(\g_aJ)_{\b\c}}\hskip-2.5cm
	R_{abcd\a}=0\komma\cr
	&R_{a\b\c}=-2ie^{-{\phi\/2}}(\g_aJ)_{\b\c}\komma
	\phantom{R_{abc\a\b}=-2i(\g_{abc}I)_{\a\b}}\hskip-2.5cm
	R_{ab\c}=-e^{-{\phi\/2}}(\g_{ab}J\LL)_{\c}\komma\cr
	&R_{\a}=2e^{-\phi}(I\LL_{\a})\punkt\cr}
\Eqn\rrconfig
$$
All curvatures with more than two spinor indices have dimension lower
than zero, and must vanish.
We now insert these constraints in $\d_{\k}\L_{\wz}$ and at the same time
substitute $\k$ by $\G\k$, 
where the projection matrix in a general background is
$$
\G={\e^{ijkl}\/\sqrt{-\det(\hat g+e^{-{\phi\/2}}\hat\F)}}
	\left(\fraction{24}\g_{ijkl}I
	-\fraction{4} e^{-{\phi\/2}}\hat\F_{ij}\g_{kl}J+
	\fraction{8}e^{-\phi}\hat\F_{ij}\hat\F_{kl}I\right)\punkt\Eqn\pgen
$$
Note that the denominator is just minus the DBI lagrangian.
It is then just a matter of writing out the various terms and verifying
that they cancel the ones coming from the variation of the DBI term, thus  
proving the $\k$-symmetry of the action $I=I_{\dbi}+I_{\wz}$. Since the
constraints that go into this calculation can not be relaxed 
while keeping the system $\k$-invariant\footnote{*}{Although the constraints 
given in equations ({\xold3}.{\xold8}) and ({\xold3}.{\xold13}) 
depend to a certain extent
on the choice of conventions, this choice is physically irrelevant.}, 
we have also shown
the equivalence between demanding $\k$-invariance and imposing the 
field equations for a D3-brane propagating in a type IIB supergravity
background.

Here we make the observation that the projection matrix in (\pgen) can be 
expressed in a more elegant way as 
$$
\G\,d^4\xi = -{1\over\L_{\dbi}}\exp\left(e^{-\phi/2}{\hat\F}\right)\g I\komma
\Eqn\elegantgamma
$$
where  
$$
\g \equiv \bigoplus_{i\in{\Bbb N}} \g_{(2i)}\,(-K)^i 
	= 1\!\!1 - \g_{(2)}K + \g_{(4)}\komma\Eqn\smallgamma
$$
and it is understood that only terms of form degree four are to be 
included in the
expansion. Considering that the WZ term takes the universal form (\superWZ)
and that this form bears a close resemblance to (\elegantgamma), 
we believe that the latter expression may be relevant
for a general formulation of supersymmetric D-branes.

To make this picture complete, we would like to establish contact with
the type \II B supergravity formulation of ref\.\ [\HoweWest].
The torsion components at dimension $0$ and ${1\/2}$ are
stated in equation (\nsnsconfig) and the Bianchi identities for the torsion
are straightforward to check at this level. 
The constrained components of the SL$(2;\R)$-invariant 
three-form field strengths are
$$
\left[\matrix{\Ht\cr\Rt\cr}\right]_{\a\b c}
	=\left[\matrix{2i(\g_cK)_{\a\b}\cr -2i(\g_cJ)_{\a\b}\cr}\right]
\komma\qquad
\left[\matrix{\Ht\cr\Rt\cr}\right]_{\a bc}
	=\left[\matrix{-(\g_{bc}K\LL)_\a\cr-(\g_{bc}J\LL)_\a\cr}
	\right]\komma\eqn
$$
and they fulfill the Bianchi identities
$$
D\left[\matrix{\Ht\cr\Rt\cr}\right]
	+M\left[\matrix{\Ht\cr\Rt\cr}\right]=0\komma\eqn
$$
where $M$ is the connection formed from the scalar field 
$2\kross2$ matrix $\V$ 
belonging to the group manifold of SL$(2;\R)$: 
$$
M=\V^{-1}d\V=\left[\matrix{P&P'+Q\cr P'-Q&-P\cr}\right]\punkt\eqn
$$
The physical scalars belong to
the coset SL$(2;\R)/\,$U$(1)$, and the covariant derivative includes the 
U$(1)$ part, $Q$, of the connection. The SL$(2;\R)$-invariant three-form 
field strengths
carry U$(1)$ charge two, so the dimension zero components have charge zero.
It is consistent with the Maurer--Cartan equation for $M$,
$$
\eqalign{&0=dP+2Q\w P'=DP\komma\cr
	&0=dP'-2Q\w P=DP'\komma\cr
	&0=dQ+2P\w P'\komma\cr}\eqn
$$
to make the gauge choice $Q\=-P'$, thus making the connection
lower-triangular, and this allows one to solve the Maurer--Cartan 
equations for $P$ and $P'$ in terms of the physical scalars 
as $P\={1\/2}d\phi$, $P'\={1\/2}e^\phi dC_0$. 
We have $P'_\a\=(I\LL)_\a$.
After rescaling $\Ht$ with $\exp{({\phi\/2})}$ and $\Rt$ with
$\exp{(-{\phi\/2})}$ one arrives exactly at the configuration of equations
(\nsnsconfig) and (\rrconfig)
and the Bianchi identities (\bianchiidentities). 
When checking the
Bianchi identities $dH\=0$, $dR_{(3)}\+H\w R_{(1)}\=0$
(which are absolutely necessary, considering the way the potentials
enter the action above) explicitly, we need to
observe that these field strengths carry U$(1)$ charge zero, so the
dimension zero components are of charge $\!-2$. It is a convincing test of
the procedure carried out above that it actually dictates the constraints
and Bianchi identities for the fields of type \II B supergravity.

Having established the $\k$-symmetry of the D3-brane action in an
on-shell type IIB supergravity background, it is instructive to return
to the case of flat superspace. As mentioned in section {\old2}, the flat limit
should reproduce all the $\t$-extensions introduced in the beginning
of that section. From the only non-vanishing component of the flat torsion, 
${T_{\a\b}}^c\=2i(\g^c)_{\a\b}$,
and the definition of the superspace torsion we conclude that one choice of
flat vielbeins is
$$\eqalign{
&{E_m}^a={\d_m}^a\komma
\phantom{{E_{\mu}}^a=i{\d_{\mu}}^{\a}(\g^a \t)_{\a}}\hskip-1cm
{E_m}^{\a}=0\komma\cr
&{E_{\mu}}^a=i{\d_{\mu}}^{\a}(\g^a \t)_{\a}\komma 
\phantom{{E_m}^a={\d_m}^a}\hskip-1cm
{E_{\mu}}^{\a}={\d_{\mu}}^{\a}\punkt\cr}\eqn
$$
This immediately leads to the identification of the one-form $\P^a$
used in section {\old2} with the flat limit of 
$$
(dZ^M){E_M}^a=dX^m{\d_m}^a+id\bar\t\g^a\t=\P^a\punkt\eqn
$$

In the case of antisymmetric tensor fields we would like to solve
the flat constraints on the field strengths 
for the corresponding flat potential. The definition of the flat
NS-NS three-form $H_{ABC}$ in terms of the flat torsion is
$$
H_{ABC}=3\*_{[A}B_{BC)} + 3{T_{[AB}}^D B_{|D|C)}\punkt\eqn
$$
Here $[\ldots)$ refers to graded symmetrizations.
The only non-zero component of $H_{ABC}$ in the flat limit is 
$H_{a\b\c}\=2i(\g_aK)_{\b\c}$,
which is reproduced using the potential
$$B_{ab}=0\komma\qquad B_{a\b}=-i(\g_aK\t)_{\b}\komma\qquad
	B_{\a\b}=-(\g^a\t)_{(\a}(\g_a K\t)_{\b)}\punkt
\eqn
$$
Thus, we find that $\F_{ij}\=F_{ij}\-{E_j}^B{E_i}^A B_{AB}$ 
gives exactly the $\t$-extended $F$ used in section {\old2}. 
The other field strengths work in a similar way.

\section\comments{Conclusions and comments}In this paper we have so far 
shown that the sum of the DBI and the WZ terms
that represents the coupling of the D3-brane to a background four-form
potential of type IIB supergravity can be made both $\k$-symmetric 
and supersymmetric in target space in a unique and rather straightforward way. 
In ref\.\ [\CGMNW] the bosonic part of the DBI action for the D3-brane 
was shown to be equivalent on-shell to the BDHT-type action 
$$
\L=\sqrt{-\g}\left\{-\half \tr(\g^{-1}g)+
	\sqrt{(1+\fraction{4} \tr(\g^{-1}\hat F)^2)^2
	+\D(\g^{-1}\hat F)}\,\,\right\}\eqn
$$
by using the field equation for the world-volume metric $\g$ to algebraically
eliminate it. $\D$ is given by $\D(X)\=\det X\-{1\/16}(trX^2)^2$. This kind of 
bosonic action was discussed recently for several important cases by 
Townsend [\Townsend], and by Tseytlin [\Tseytlin] who demonstrated that
together with the WZ term the bosonic action exhibits certain duality
properties.
Furthermore, the square root in the above action can be removed
by introducing also a non-propagating scalar on the world-volume:  
$$
\L=\sqrt{-\g}\left\{-\half \tr(\g^{-1}g)+
\varphi\Bigl[\bigl(1+\fraction{4} \tr(\g^{-1}\hat F)^2\bigr)^2
+\D(\g^{-1}\hat F)\Bigr]
+\varphi^{-1}\right\}\punkt\eqn
$$
The $\k$-symmetrization of this action follows the same pattern
as for the DBI action. In fact, one can easily convince oneself that 
the solution presented here for
the transformation rules work equally well for the BDHT form of the 
action. It is interesting to note in this context that using the so-called
1.5 order formalism when establishing the invariance of the action, it
does not matter whether the matrix $\G$ in the projection matrix contains 
the super-DBI action
in the denominator or the superversion of the BDHT-type action derived in
ref\.\ [\CGMNW]; in any case, the denominator will not be as 
simple as $\sqrt{-\g}$,
and the analysis of ref\.\ [\CGMNW] makes it quite clear that the 
cases $p\=1,2$ are special in this respect. 

Our results may thus be summarized as follows: the action
$$
\eqalign{I_{\hbox{\fiverm D3}}=&\int \sqrt{-\g}\left\{-\half \tr(\g^{-1}\hat g)
	+\varphi\Bigl[\bigl(1+\fraction{4} \tr(\g^{-1}
		e^{-{\phi\/2}}\hat \F)^2\bigr)^2
	+\D(\g^{-1}e^{-{\phi\/2}}\hat \F)\Bigr]
	+\varphi^{-1}\right\}\cr
	 +&\int e^{\hat \F}\hat C \cr}\eqn
$$
and all other forms of it obtained by using algebraic world-volume 
field equations are invariant under the $\k$-transformations
of section {\old3}, provided the background fields satisfy the IIB supergravity
field equations. 

We have not explicitly discussed the properties of our D3-brane action
under the SL$(2;\Z)$ S-duality group (which is the quantum remainder
of the SL$(2;\R)$ above) of the type \II B string theory.
Knowing how the self-duality works in the bosonic case [\Tseytlin,\Gutperle], 
it should be fairly easy to repeat those calculations with all background
fields replaced by the corresponding superfields.

It would also be interesting to study the system under consideration in terms
of world-volume instead of space-time supersymmetry. The field content matches
that of $N\=4$ super-Maxwell theory 
(or super-Yang--Mills in the non-abelian case): 
a gauge potential, four spinors (we envisage that gauge fixing the
$\k$- and reparametrization symmetries will turn the world-volume 
scalars $\t$ into world-volume
spinors, as happens for the fermions in the Green--Schwarz superstring),
and six scalars (the transverse components of $X$). The DBI action belongs
to a very restricted class of non-linear 
generalizations of the Maxwell theory that upon supersymmetrization does not
give rise to propagating auxiliary fields [\Karlhede], and it is most likely
that the maximally extended world-volume theory, should it be 
quantized, is a finite quantum field theory. 
It is not obvious that there will be reason to perform
such a quantization in a string theory context --- being D-branes, these
objects do not seem to play the r\^ole of truly fundamental constituents
in any ``string theory'' vacuum. At the present level of understanding,
the actions treated in this paper play a r\^ole as low-energy effective
actions, whose ``classical'' variation gives the necessary correction to
the open string $\b$-function. We should not be surprised, however, if a
more profound understanding of non-perturbative ``string theory'' would 
involve something like quantized D-branes. 

While this paper deals exclusively with 3-branes, it is clear
that the mechanisms at work for other type \II A or \II B Dirichlet branes
will be quite analogous, and once the ice is broken the formulation
of the corresponding actions is a matter of algebra.
We also believe, although this is less obvious, that the additional insights
gained here may be used in a broader context, namely for $p$-brane 
actions with antisymmetric tensor supermultiplets 
[\Townsend]. Interesting
cases that fall into this category are the five-brane in eleven dimensions
[\HoweSezgin,\BdRO] 
and perhaps also $p$-branes living in target spaces of dimension twelve and
higher [\Hull,\Vafa,\Bars,\Martinec,\Townsend]. 
The generalization of the DBI action
to non-abelian vector fields [\WittenIII]
is another obvious issue. Unfortunately, 
in spite of the fact that there are many possible non-abelian generalizations 
of the BDHT-type actions given above, not much can be achieved until
one understands the technicalities of how to eliminate the world-volume 
metric from the action. We hope to come back to these questions in the future.
 
\frenchspacing
\refout
\end